\documentstyle[11pt,newpasp,twoside,epsfig]{article}
\def\edcomment#1{\iffalse\marginpar{\raggedright\sl#1\/}\else\relax\fi}
\reversemarginpar

\begin{document}
\title{The Magnetic Fields of the Universe and Their Origin}
 \author{Stirling A. Colgate \& Hui Li}
\affil{Los Alamos National Lab, T-6, MS B288, Los Alamos, NM 87545 }

\begin{abstract}
Recent rotation measure observations of a dozen or so 
galaxy clusters have revealed a surprisingly large amount of 
magnetic fields, whose estimated energy and flux are, on average, $\sim
10^{58}$ ergs and $\sim 10^{41}$ G cm$^2$, respectively. 
These quantities are so much larger 
than any coherent sums of individual galaxies within the cluster 
that an efficient galactic dynamo is required.
We associate these fields with single AGNs within the cluster
and therefore with all galaxies during their AGN phase. Only the
central, massive black hole (BH) has the necessary binding energy,
$\sim 10^{61}$ ergs.  Only the accretion disk during the BH formation
has the winding number, $\sim 10^{11}$ turns, necessary to make the
gain and magnetic flux.  We present a model of the BH accretion disk
dynamo that might create these magnetic fields, where the 
helicity of the $\alpha - \Omega$ dynamo is driven by
star-disk collisions. The back reaction of the saturated dynamo forms
a force-free field helix that carries the energy and flux of the
dynamo and redistributes them within the clusters.
\end{abstract}

\section{Introduction}

The problem of understanding the origin of the large scale galactic
magnetic fields has been with us for over forty years.  
There have been many papers and reviews on the galactic
and extragalactic magnetic fields 
(see Moffatt 1978; Parker 1979; Krause and Radler 1980; 
Ruzmaikin et al. 1988; Wielebinski \& Krause 1993;  Beck et al. 1996; 
Zweibel \& Heiles 1997; Kulsrud 1999),
and  observational reviews (see Miley 1980; Bridle \& Perley 
1984; Kr\"onberg 1994), including the observations themselves
(e.g. Perley et al. 1984; Taylor et al. 1990; Taylor \& Perley
1993; Eilek et al. 1984).

Recent rapid progress in observational work on galaxy clusters
has revealed a surprising result. Intracluster medium (ICM) appears
to be definitely magnetized, and in many cases, perhaps are highly
magnetized as convincingly argued by Eilek et al. (2000).
Figure 1 presents one such example in Hydra A Cluster as shown
by the rotation measure ($R_m$) map made by Taylor \& Perley (1993).
We will show in this article that the implied magnetic energy
and flux estimated from extensive $R_m$ maps of a dozen or so 
galaxy clusters are so exceedingly large that the conventional
galactic dynamo models may prove to be inadequate. We argue that
a new source of energy and a different form of the galactic dynamo
are required.

As the rotation measure observations of galaxy clusters are relatively 
new and some of them are (yet) unpublished by the observation teams, 
we will first explain some of the observation results in detail, then
discuss their physical implications at length. In the second half
of the article, we will propose a new paradigm related to AGN
accretion disks and describe some of our recent efforts in understanding
a sequence of physical processes revolving around the origin of
cluster magnetic fields. 

\section{Galactic and Extragalactic Magnetic Fields}

Faraday rotation measures, $R_m$, are shown to be consistent with six
other physical interpretations of magnetic fields in ours and
nearby galaxies (star light polarization, interstellar Zeeman
splitting, synchrotron emission, synchrotron polarization, and
inferred by x-ray emission and cosmic ray isotropy and
pressure) (see Kr\"onberg 1994 for a review), thus establishing
$R_{m}$ as a reliable measure of galactic and extra galactic
magnetic fields. Because of the existence of many
self-illuminating as well as background sources, usually AGN,
and the increasing sensitivity of radio detection,
$R_{m}$ has become the recognized measure of  extragalactic
magnetic fields (Kr\"onberg 1994; Taylor et al. 1994;
Ge \& Owen 1994; Krause \& Beck 1998). 

\subsection{Magnetic Flux and Energy in Galaxy Clusters}

Recently, high quality $R_{m}$ maps of self-illuminating sources of galaxy 
clusters where the distances are known have become available
(for example, Taylor \& Perley 1993; Eilek et al. 2000).
An important quantity that has received less discussion in these
papers is {\it the magnitude of the magnetic flux and energy}. 

Figure 1 shows the $R_m$ map of the region illuminated by 
Hydra A in the cluster (courtesy of Taylor \& Perley 1993). 
The largest single region of highest field in this map
has approximately the following properties: the size
$L \simeq 50$ kpc and $B \simeq 33 \mu$ G, derived on the basis 
that the field is patchy and is tangled on a 4 kpc scale. This leads
to a startling estimates of flux, $\it{F} \approx BL^2
\simeq 8 \times 10^4 \mu$ G $kpc^2$, and energy, $W = (B^2/8\pi) L^3
\simeq 4 \times  10^{59}$ ergs, assuming that the tangled field is only 
confined to the 50 kpc region. If this is extended to the whole
cluster which is $\sim 500$ kpc,
then the implied flux and energy are correspondingly larger by a
factor of $100$ and $10^3$, respectively.
A similar conclusion can be reached when a larger sample of $R_m$ of
galaxy clusters are analyzed using the data presented in 
Eilek et al. (2000). In Table 1, 
we have reproduced  part of table given in Eilek et al. (2000)
and added two columns where the
approximate flux and energy are  calculated assuming that the fields
are partially tangled or in loops.  

\begin{table}
\caption{A list of cluster core parameters and their estimated
magnetic fluxes and energies. Mean magnetic field is taken to be
$\sim \sqrt{3}\times \langle B_{\parallel}\rangle$.
Data in the first three columns are
taken from Eilek et al. (2000).}
\begin{tabular}{|l|c|c|c|c|}
\tableline
Source & Size  & $\langle B_{\parallel}\rangle$ &
$\langle B^2 L^3 / 8\pi \rangle$ & $\langle BL^2
\rangle$ \\
& (kpc) & ($\mu$G) & ($10^{58}$ ergs) & ($10^{41}$G cm$^2$) \\
\tableline 
A 400 & 100 & 2.9 & 3 & 5 \\
A 1795 & 7 & 18 & 0.03 & 0.17 \\
A 2052 & 8 & 17 & 0.05 & 0.17 \\
A 2029 & 10 & 1 & 0.0003 & 0.02 \\
A 2199 & 30 & 15 & 2 & 0.3 \\
A 2634 & 140 & 1.9 & 30 & 7 \\
A 4059 & 10 & 69 & 0.15 & 1\\
Cyg A & 70 & 15 & 25 & 1.5 \\
Hydra A & 50 & 33 & 40 & 15\\
Virgo A & 3 & 35 & 0.01 & 0.05\\
\tableline 
\end{tabular}
\end{table}

Furthermore, the estimated values of fluxes and energies are most
likely to be the minimum of the actual magnetic fields existing in
the galaxy clusters. 
Faraday rotation depends upon the component of the field
strength along the line of sight,
$B_{\parallel}$, the distance along the line of sight,
$Z_{o}$, and electron density, $n_{e}$. 
Estimates of  $n_{e}$ can be made from the x-ray
emission measurements of the clusters with a typical accuracy of
$\sim 20\%$, and it varies by  factors of 2 to 4  over the region of
the source, but otherwise is nearly uniform, and  clumping is
small (Taylor et al. 1994).   If the field
is folded in any fashion so that regions of oppositely directed
field are in the line of sight, then the observed
$R_{m}$ will be smaller than that if the same field lines  were
straightened out  into one direction.  In other words
$R_{m}$ is a minimum measure of $B_{\parallel}$.

To put the above numbers in perspective, for a typical galaxy like ours,
e.g., with 1 kpc thickness, 3 kpc Homberg radius, and a field of
$\sim 3 \mu$ G, the magnetic flux and energy are roughly  
$10^{38}$ G cm$^2$ and $4\times 10^{52}$ ergs, respectively.
One observes that the
flux and energy given in Table 1 range
from close to the Hydra A limit  to no more than
$10^2$ times that of a typical galaxy.

The magnitude of the implied fluxes
and energies are so large, $\times 10^3$ and $\times 10^6$
respectively, compared to these quantities within standard galaxies
that their origin requires a new source of energy and a different form
of the  dynamo than previous  galactic models. These minimum energies
are sometimes even larger than the baryonic binding energy of galaxies
($\sim 2 \times 10^{58}$ ergs). The extremely large fluxes also seem 
out of reach via amplification by 
ordinary galaxy rotations in a Hubble time.

Next, we discuss the difficulty with using
turbulence to create these  nearly uniform, highly correlated
and coherent regions of
$R_{m}$ as seen in Figure 1. We then discuss the still greater 
difficulty of creating the total magnetic energy of the cluster  based upon
a turbulence dynamo model.

\subsection{Turbulent versus Coherent Fields}

It has been suggested by a number of people 
(Eilek 1999;  DeYoung 1980; Ruzmaikin et al. 1989;  Goldman \& Rephaeli
1991; Jaffe 1980) that the entire cluster is uniformly
turbulent due to Rayleigh Taylor instabilities during matter
in-fall into the cluster,  and that this turbulence drives the
cluster dynamo creating the fields.  The problem with this
interpretation is the total  magnetic energy, the magnitude of
the turbulence, the strength of the fields, the apparent
correlation of
$R_{m}$ maps with single AGN structures, and finally the
limited number of rotations of the cluster in a Hubble age.

Because of the small rotation rate of the typical cluster,
$\sim 100$ km $s^{-1}$, the available rotation energy is
small, $\sim 10^{-2}$ of the cluster binding energy, which
has a thermal velocity of $\sim 10^3$ km $s^{-1}$. So applying
the turbulent model to Hydra A implies a magnetic energy
$10^3$ greater than the rotational energy.  Therefore
the dynamo must be of the  $\alpha^2$ type  where fields are
generated on the small scale, yet as Taylor et al. (1994) point
out, the fields of Hydra A and A1795 reverse on the different sides of the
core, requiring coherence on scales of 100 kpc. Since this
reversal is correlated with the structure of the source, and since
the energy generated at the small scale is small compared to
the turbulence input and the turbulent input should be small
compared to the binding energy (DeYoung 1992), we believe that all
these factors point to  random, localized sources of
magnetic energy of size $> 10^{60}$ ergs. This is probably too
demanding for turbulence.

Furthermore, it will be difficult to produce large scale
coherent $R_m$ regions, which have been observed in Hydra A (Figure 1,
northern region) and several other galaxy clusters (Eilek et
al. 2000). This is because in a  turbulent
plasma, the emission, the $R_{m}$, and the degree  of polarization
should all be statistically  symmetric. Despite the
unlikelyhood of all these factors  conspiring to create both a
pattern and a nearly uniform
$R_{m}$, one observes in  many $R_{m}$  maps of AGN, mostly in
clusters, a distinct match in the $R_{m}$ pattern with the jet
like pattern of emission. Particularly the sign of the average
$R_{m}$ in several cases  reverses across a symmetry plane
through he core of the AGN (Taylor \& Perley 1993).  The size
of the regions of uniform
$R_{m}$ correlates strikingly with the size of the jet as a
function of distance from the nucleus. We interpret this
correlation as due to  the source of the field being the AGN
jet as opposed  to a turbulent
$\alpha - \Omega$ dynamo in the cluster  as a whole.

\subsection{Average Field Structure}

Using serendipitous polarized background sources,
therefore random lines of sight through random clusters,
Clark  \& Kr\"onberg's (1999) have made $\sim 80$
$R_{m}$ measurements. Their observations
have produced a boundary of the typical cluster in
$R_{m}$ such that the average field was $\sim 3 \mu$ G out to a
radius of
$R_{cluster} \simeq 300$ kpc.  The magnetic flux and energy,
$10^{4}\mu$ G $kpc^2$ and $10^{60}$ ergs, are then similar to the
largest structure already discussed in Hydra A.  If each galaxy 
of a typical cluster with $\simeq 50$ large galaxies contributes a
high field region during its AGN phase, then the probability of
intersecting such a region of area that is $\simeq 1\%$ of the cluster 
is roughly $\simeq 5 \times 10^{-3}$ so that in 100 lines of sight, the
probability of intersecting a Hydra-like region of an AGN in a
cluster is $\simeq 50\%$. This is not inconsistent with the variability
they observed.  Finally we note that the large degree of
polarization observed $\sim 50\%$ in these sources indicates
that  the rotation source and emission source cannot be in the
same location (Burn 1966; Taylor 1991), otherwise polarized
emission from various depths in the source would undergo
different degrees of rotation and hence emerge depolarized.
Therefore in any model the Faraday screen and the emission
source must be related and even congruent in order that the
screen and hence $R_{m}$ be correlated with the core of the
AGN.

\subsection{Black Hole Accretion Disk as the Engine}

Purely based on the energetics, the accretion disk
around supermassive black holes in AGNs offers an attractive
site for the production of magnetic fields.
The accessible binding
energy of the black hole is $\sim 10^8 M_{\odot} c^2 \sim 10^{61}$
ergs and the winding number of the disk forming the BH of 
nearly every galaxy is $N_{w} \sim 5 \times 10^{10}$ at $10R_{g}$, 
where $R_{g}$ is the BH horizon ($\sim 1$AU). Using the canonical
numbers thought to apply to AGN disks,  the BH dynamo flux can be
$\it{F_{\rm BHdyn}} \simeq B_{\rm BHdyn}\pi R_{\rm BHdyn}^2 N_{w}
\simeq 10^{43}$ G cm$^2$, where we have used
$B_{\rm BHdyn}\simeq 10^4$ G at $L_{AGN} \sim 10^{46}$ ergs $s^{-1}$  
and $R_{\rm BHdyn} \simeq 10 R_{g} \approx 10^{14}$ cm. 
Both the flux and energy  from this
simple analysis are $\sim 10$ times the maximum observed values.  No other
source of energy is likely to be sufficient by many orders of
magnitude. Therefore it is  much more reasonable to assume that every
AGN, both within and external to clusters, produces the
magnetic energy and flux that we observe in this extreme case
from the binding energy released in the accretion disk forming
the central  BH.  This implies that every galaxy contains a BH where
$\sim 90\% - 95\%$ of the accessible binding energy is
transformed into magnetic energy during its AGN phase by an accretion
disk dynamo.  On the average this flux and
energy is distributed  throughout the universe as
force-free fields  and only a small fraction  $5\%- 10\%$ of the
magnetic energy is dissipated  in the form of the AGN spectra, thus
explaining the problem of the  missing AGN  luminosity (Richstone 1998;
Krolik 1999).   In this
picture  a larger fraction  of the magnetic energy is
dissipated where the brightest AGNs are seen in galaxy clusters,
because only in the clusters is a sufficient gas density retained by
the gravity of the cluster such that this density confines the  field
increasing the fraction  of
the magnetic energy that is dissipated.  For
most galaxies external to dense clusters a small fraction of this
magnetic energy is dissipated as the AGN radiation, a small  fraction
remains in the galaxy, and the bulk of the magnetic energy and flux  is
distributed in the walls and voids of the universe.

\section{Astrophysical Requirements and Progress with a Model}

The sequence of phenomena that can explain this astonishing
extragalactic magnetic flux and energy  must start with an accretion disk
forming a massive central galactic  BH. This in turn presumes an
answer to an equally enigmatic question, namely the formation of these
massive galactic BHs themselves (Begelman et al. 1989;
Rees 1999). By focusing on the transport of angular momentum we
believe that the flat rotation curve mass distribution
can be explained as a plausible result of any non-linear
collapse of an initial gaseous baryonic density  fluctuation by
hierarchical  tidal torquing (Newman \& Wasserman 1999). The BH forms
from this mass distribution  when the Rossby vortex torque mechanism
supersedes tidal torquing and an accretion disk forms.  All this mass
then collapses to a BH.  The flat rotation curve,
$M \propto R$, results in
$\Sigma \propto R^{-1}$.   When this thickness  reaches  $\Sigma \simeq
100$ to $1000$ g
$cm^{-2}$, heat is confined for several revolutions,  and
the Rossby vortex instability initiates at $M_{disk} \sim 10^7$ to
$10^8 M_{\odot}$.   Finally the dynamo produced fields then supersede
the previous torque mechanisms.

\subsection{The Rossby Vortex Torque Mechanism}

We have predicted and  demonstrated analytically and
numerically how a new instability in
Keplerian flow, the Rossby vortex instability, can grow
(Lovelace et al. 1999; Li et al. 2000a; Li et al. 2000b). The
production of vortices is shown in Figure 2. This
instability  produces  torque and thus transports angular momentum
within an accretion disk by purely hydrodynamically means via the
interaction of large, two-dimensional, co-rotating  Rossby vortices.
The enhanced transport of angular
momentum by co-rotating vortices is recognized
in  rotational atmospheric flows (Staley \& Gall 1979)
and  in laboratory measurements of the
Ranque-Hilsch tube (Hilsch 1946; Fr\"ohlingsdorf
\& Unger 1999, Colgate \& Buchler 1999).

\subsection{The Dynamo, Star-Disk Collisions, and Helicity}

A coherent  dynamo can form in a Keplerian accretion disk
because of the large azimuthal velocity shear,
provided that there exists a robust source of non-axisymmetric
helicity.  Classically turbulence has been invoked to explain
this helicity using the mean field dynamo theory, but we know of no
way to create this degree of turbulence, with vertical
motions,  hydrodynamically  in an accretion disk, because
hydrodynamic turbulence alone is damped in an accretion disk
(Balbus \& Hawley 1998). The magnetic instability of Balbus
\& Hawley  will lead to turbulence, but the magnitude of the
turbulence is orders of magnitude too small compared to the
Keplerian stress. Instead we have identified  a new, robust
source of helicity driven by star-disk collisions by a small
mass fraction
$\sim 10^{-3} - 10^{-4}$ of pre galaxy-formation stars. The Keplerian
shear and a star-disk collision with the  twist producing helicity is
shown in Fig. 3.  We have demonstrated by laboratory flow
visualization experiments of how plumes, driven  in a rotating frame,
counter rotate relative to the frame (Beckley \& Colgate 1998;
Beckley et al. 2000) and thus produce a robust and coherent helicity
where flux is always added in the same direction and where the driving
force is large compared to the Keplerian stress in the disk.

We  have simulated the positive, exponential gain of both the
quadrupole and dipole poloidal field of such a dynamo with a
vector potential code in 3-D, cylindrical coordinates, where
the velocity field simulates both the Keplerian rotation and
star collision-produced plumes.  We have
observed a growth rate of $\sim 10\%$ per revolution, two
plumes per two revolutions, $R_{plume} = 1/3 R_{disk}$, and with a
magnetic  Reynolds number, $R_{ey,\Omega,B} =  100$   (Pariev,
Colgate \& Finn 2000).

\subsection{The Saturation of the  Dynamo and the Formation of the
Helix}

With positive gain and large winding number, the
dynamo will saturate regardless of how small the seed field is.  Since the
helicity does not depend on turbulence, it will not be subject to
turbulent
$\alpha$-quenching at the small scale (Vainstein \& Cattanio
1992;  Vainshtein \& Rosner 1991).  Furthermore since the
stars maintain virial velocity, their velocity is supersonic
relative to the disk and the resulting shock stress is large. At the
back reaction limit, the field grows until the torque of the field
affects the  Keplerian motion, and the accessible BH binding energy  is
converted into magnetic energy.    The progressive
loss of this flux is a force-free helical, Poynting magnetic
flux, which we identify as the collimated AGN jets.   We have investigated
the field topology of these twisted
helical flux surfaces by integrating the Grad-Shafranov
equations for a force-free axisymmetric  field  with a
Keplerian distribution of winding number (Li et al. 2000c)
as shown in Fig. 4.  Since the field decreases as $B_{helix}
\propto 1/R$, the pressure, at large radius as the helix extends to
Mpcs, becomes of the  order of the IGM and and the outer boundary of
the helix is self collimating (Lynden-Bell 1996).
The energy carried by this helix, at a mean
radius near the BH,
$R_{dyn}
\simeq 10R_{BH}$ is the   accessible energy of accretion or
$\dot{M}_{BH} c^2/10 = (B_{helix}^2/8
\pi) (100  \pi R_{BH}^2)$ or $B_{helix} \simeq 10^4$ G, $I =
5R_{helix}  B_{helix} = 5  \times  10^{18}$ amperes, and
$V_{potential} = 10^{20}$ volts and $I \times V_{potential} =
10^{39}$ watts $= 10^{46}$ ergs $s^{-1}$.  General relativity inside
the innermost stable orbit will add additional energy (Blandford \&
Znajek 1977; Livio et al. 1999).

\subsection{$J_{\parallel}$ Reconnection and Acceleration}

The distribution of this flux in the universe occurs by partial
tearing mode reconnection producing the minimum energy Taylor state
(Taylor 1986).  The total flux is conserved, but a fraction of the
energy is dissipated in the tearing mode,  $J_{\parallel}$
reconnection.  The resulting
$E_{\parallel}$ acceleration of the current carriers produces the
emission that we associate with the  AGN.

\begin{figure}
\end{figure}

\acknowledgments
We are indebted to Richard Lovelace, Howard
Beckley, Vladimir Pariev,
John Finn, Mike Warren, Dave Westpfahl, Van Romero, Ragnar
Ferrel,  and Warner Miller for  direct contributions to this project
and to very many more who have contributed in discussions, criticisms
and encouragements.  HL acknowledges the support of an Oppenheimer 
Fellowship. This research is supported by the DOE,
under contract W-7405-ENG-36.

\end{document}